# (110) Facet of MgTe Zinc Blende Semiconductor: A Holy Grail for Modern Spintronics

Manish Kumar Mohanta* and Puru Jena[†]

Department of Physics, Virginia Commonwealth University, Richmond, VA 23284, USA

E-mail: *manishkmr484@gmail.com, mohantamk@vcu.edu, [†]pjena@vcu.edu

Unlike, momentum-dependent Rashba spin-splitting, materials exhibiting intrinsic momentum-independent unidirectional spin polarization also known as persistent spin texture (PST) in the full Brillouin zone are scarce. In this work, a list of characteristic electronic properties for identifying an ideal PST material is provided based on earlier analytical models, and a new semiconductor, the MgTe(110) facet is proposed which satisfies all these conditions and exhibits PST in the full Brillouin zone. The atomic arrangement in this particular facet exhibits three basic symmetries found in nature: rotation, reflection, and translation. Using the method of invariance, an effective Hamiltonian is constructed which reproduces the results obtained using the density functional theory. Further, mono/few layers of MgTe (110) facets of the zinc-blende structure are proposed for a ferromagnet-free non-ballistic spin-field effect transistor (s-FET) that combines both the spin-Hall effect and inverse spin-Hall effect, thus harmonizing spintronics with conventional electronics. Although only quantum well structures have been experimentally studied for nonballistic s-FET under the stringent condition of equal Rashba and Dresselhaus strength, PST originating intrinsically in the proposed 2D structures makes them an ideal alternate.

# 1. Introduction

The precise control of spin degrees of freedom for data storage and process has been of great interest and a hot topic of research following the proposal of a spin-field-effect transistor (s-FET) by Datta and Das. [1–3] The s-FET device consists of a lateral semiconducting channel exhibiting strong spin-orbit-coupling (SOC) and two ferromagnets used for spin generation and detection. The spin transport is controlled by the gate voltage in the semiconducting region. Depending on the spin-transport, ballistic (impurity-free) and non-ballistic s-FETs have been proposed but the latter have been less explored. In ballistic s-FET, the spin direction is maintained in the channel without any scattering. The nonmagnetic Rashba semiconductors exhibit momentum-dependent Rashba spin splitting and are thus prone to impurity scattering in a non-ballistic region. However, spin-orbit interaction can be engineered to produce momentum-independent unidirectional spin configuration which is also known as persistent spin texture (PST). This has been theoretically shown in two-dimensional quantum well systems having equal Rashba and Dresselhaus strength. [4,5] Under this condition, enforced by SU(2) symmetry, spins exhibit an extraordinarily long lifetime, even in the presence of disorder or imperfection. [6] The spin-helix state in a two-dimensional electron gas system is found to be robust against D'yakonov-Perel's spin relaxation which makes Datta-Das type s-FET operable in the nonballistic transport regime. [7] For a conventional s-FET, interfacial scattering, band mismatch, ferromagnetic materials with 100% spin-polarized current, spin injection efficiency, and long spin lifetime are major challenges that hinder realizing s-FET. [8] The control of spin precession using gate voltage requires materials having large SOC for ballistic s-FET. However, relatively low SOC quantum well structures can be used in nonballistic s-FET as demonstrated by Eberle et al.. [9] Although quantum well structures have been extensively studied for nonballistic s-FET, it is necessary to explore 2D materials that exhibit intrinsic PST. [10–13]

Recently Tao and Tsymbal have proposed the existence of intrinsic PST in bulk non-symmorphic crystal structure. [14] This has generated a surge in interest in exploring 2D materials. However, there are only a handful of 2D semiconductors that have been identified to exhibit PST to date, such as MX monolayers (M: Sn, Ge, X: S/Se/Te) monolayers. [15–17] These proposed monolayers are van der Waals solids and hence odd/even effects may arise under different stacking configurations. In this work, the electronic/spintronic properties of MgTe (110) facets are thoroughly investigated using density functional theory. MgTe (110) facet is a direct band gap semiconductor with band edges located at the Γ-point and the SOC induces a unidirectional spin polarization that spans across the full Brillouin zone. In addition, being a century-old semiconductor, its experimental synthesis is well established which makes this facet very interesting and appealing to explore non-ballistic s-FET. [18,19]

# 2. Results and Discussion

## 2.1 Symmetries Associated with Atomic Arrangments in MgTe (110) Facets and Ferroelectricity

The cubic zinc-blende (ZB) structure of MgTe has many facets. A brief crystallographic description of their geometrical views is provided in Table S1 and Figure S1 in Supplemental

Material (SM). In this work, we are particularly interested in the nonsymmorphic (110) facet where the atomic arrangements show all three basic types of transformation: rotation (R), reflection (M), and translation (t). Recently discovered black phosphorene is a typical example belonging to a non-symmorphic space group. Since MgTe is a non-van der Waals solid, the focus of this work is to explore the electronic properties associated with a two-atomic thick layer (2L) which is the basic building block for the (110) facet (see Figure S1(d)). Geometric top and side views of 2L-MgTe are presented in Figure 1. The dynamical stability is confirmed by phonon dispersions plotted in Figure S2.

Considering the two-dimensional MgTe system, the crystallographic symmetry operations under which the 2D structure remains invariant are represented in eq. (1); (i) identity operation $E$; $< Ee|\{0,0,0\} >$; (ii) one 2-fold rotation (180-degree) about the $x$-axis ($C_{2x}$) followed by a translation of $\{\frac{a}{2},\frac{b}{2}\}$; $< C_{2x}|\{\frac{a}{2},\frac{b}{2},0\} >$; (iii) one mirror symmetry in $M_{zx}$; $< M_{zx}|\{0,0,0\} >$; (iv) one glide reflection $\bar{M}_{xy}$ (mirror symmetry operation $M_{xy}$ followed by translation $\{\frac{a}{2},\frac{b}{2}\}$); $<M_{xy}|\{\frac{a}{2},\frac{b}{2},0\} >$. Here, $a$ and $b$ are the lattice constants.

Two-unit cells having opposite ferroelectric polarization MgTe $+\vec{P}_x$ and MgTe $-\vec{P}_x$ are constructed to account for the effect of ferroelectric polarization on the electronic properties. A built-in ferroelectric polarization ($P$) of magnitude $0.98 \times 10^{10}$ C/m originating from the nonsymmorphic element and broken inversion symmetry exists in the crystal structure which is comparable to those in some previous reports – Bi (110) ($0.47 \times 10^{10}$ C/m) [20], CdTe and ZnTe monolayers ($2.3 \times 10^{10}$ C/m). [21] Recent experiment on SnS monolayer confirmed finite in-plane ferroelectricity at room temperature which has similar crystal structure. [22]

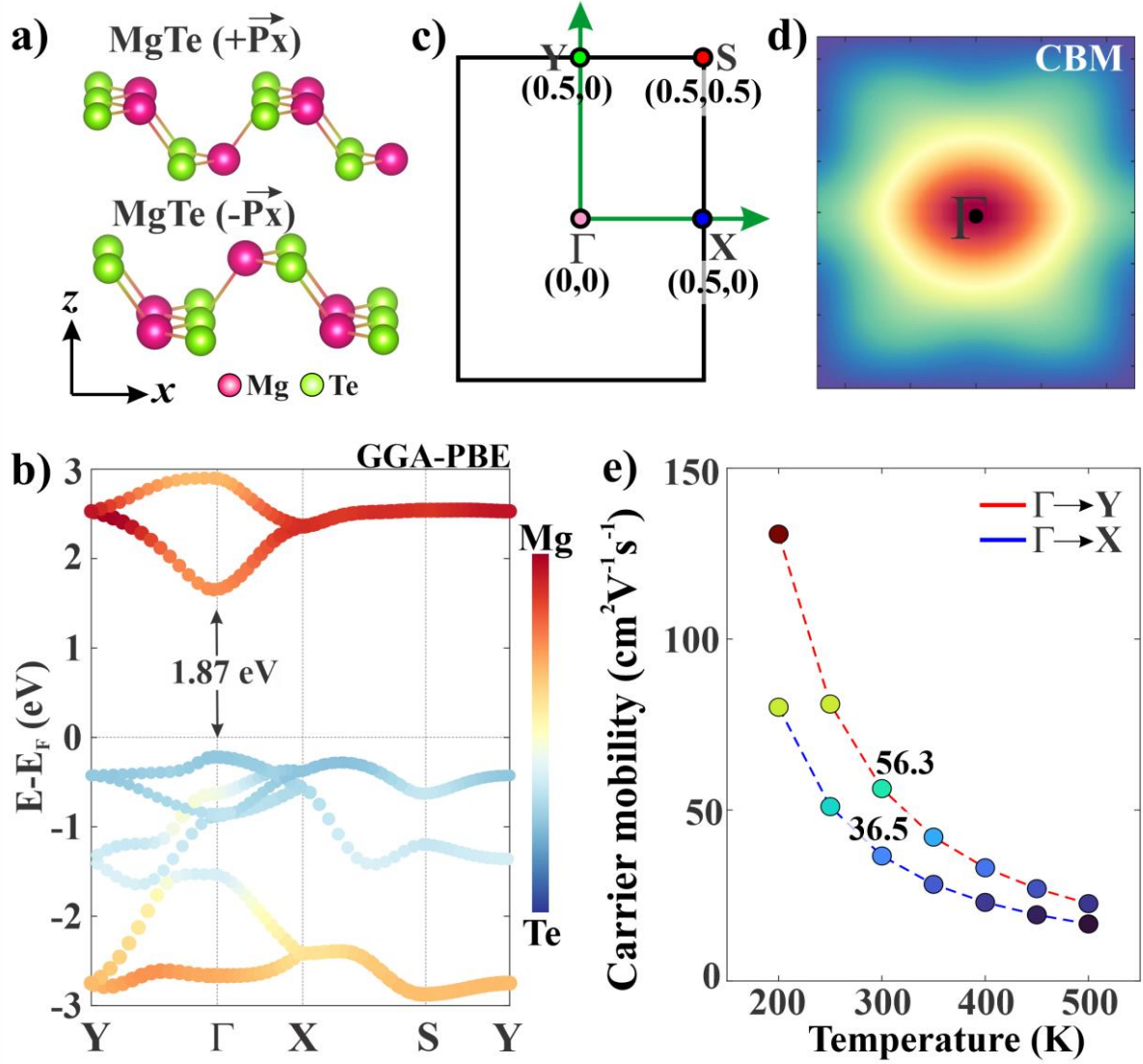

Figure 1. (a) Geometrical view of designed unit-cells having ferroelectric polarization opposite to each other, (b) atom projected electronic band structure without including relativistic effect along high symmetry points of rectangular Brillouin zone Y (0,0.5) → Γ (0,0) → X (0.5,0) → S (0.5,0.5) → Y (0,0.5), (c) schematics of rectangular Brillouin zone, (d) 2D energy contour plot of conduction band minimum (CBM), (e) charge carrier mobility as a function of temperature.

## 2.2 DFT Calculated Electronic Band Structure and Persistent Spin Texture (PST)

The electronic band dispersion of 2L MgTe is plotted in Figure 1(b) which shows MgTe to be a direct band gap semiconductor with a band gap of 1.87 eV having band edges located at the Brillouin zone centre, the Γ-point. The 2D energy contour plot of the conduction band minimum (CBM) in Fig. 1(d) indicates anisotropy in the energy dispersions along x and y direction which is reflected on the band edges of the band structure plot. The effective mass (m*) along the x-direction is larger compared to that along the y-direction which is reflected in the charge carrier mobility as plotted in Figure 1(e). The atom-projected band structure shows the contribution of each atom at the band edges. Mg and Te contribute 5.6%, 94.4% to the valence band maximum (VBM) 62.2%, and 37.8% to the conduction band minimum (CBM), respectively. Considering different ferroelectric polarization directions ($\pm P_x$ and $\pm P_y$), the band structure including SOC is plotted in Figure 2(a). The geometrical views are provided in Figure 1(a) and Figure S3. Since MgTe is composed of heavy Te-atom, the effect of SOC is reflected in Figure 2(b). In all the cases, a Rashba-type spin splitting and a spin degenerate line node (SDLN) can be observed which lifts the spin degeneracy along the line parallel to the ferroelectric polarization, respectively. For example, Rashba splitting occurs along $\Gamma \rightarrow Y$ whereas $\Gamma \rightarrow X$ remains degenerate for MgTe ($+\vec{P}_x$). The difference in constant energy Fermi surface for $\pm P_x$ and $\pm P_y$ is depicted in Figure 2(c). The Rashba constant at both band edges is calculated from the band structure using the relation [23,24] $\alpha = \frac{2E_R}{k_R}$ where $E_R$ is the energy difference between the CBM/VBM and band crossing at the $\Gamma$-point and $k_R$ is the momentum offset; $\alpha_R^{CBM}$ = 0.47 eV/Å, $\alpha_R^{VBM}$ = 1.44 eV/Å. The Rashba constant at the CBM for MgTe (110) is three times smaller compared to that in CdTe/ ZnTe (110) semiconductors (1.35 eV/Å) [21] and much smaller compared to that in GeTe (3.93 eV/Å) and SnTe (1.2 eV/Å). [15,16] Furthermore, the spin-projected band structure confirms that the band edges are contributed by an out-of-plane spin component ($S_z$) which is different from the conventional Rashba effect where the in-plane spin components ($S_x$, $S_y$) dominate at the $\Gamma$-point. [25–27] This type of spin splitting is referred to as out-of-plane Rashba spin splitting where the spins are momentum-independent and unidirectional along the $\pm z$ direction which can be depicted from the spin texture plot in Figure 2(d). A comparative plot of in-plane and out-of-plane spin projected $\langle S_x \rangle$, $\langle S_y \rangle$ and $\langle S_z \rangle$ Fermi surfaces are provided in Figure S4 for more clarity. This peculiar spin polarization in momentum space yields a specific spin-wave mode called persistent spin helix (PSH) which protects the spin from decoherence in a diffusive transport regime. This, in turn, leads to an infinite spin lifetime. [5,28] The shifting of parabolic energy dispersion $\left(\vec{Q} = \frac{2m\alpha}{\hbar^2}\hat{y}\right)$ of spin-up and spin-down is shown in spin texture plot and the pitch of the PSH $\left(l_{PSH} = \frac{2\pi}{|\vec{Q}|} = \frac{\pi\hbar^2}{m_y^*\alpha}\right)$ is calculated to be 11 nm which is approximately three times larger than that in CdTe (110) and ZnTe (110) structures and comparable to that in SnTe (001) thin films. [17] The factor of three comes from the comparative Rashba constant value as mentioned earlier.

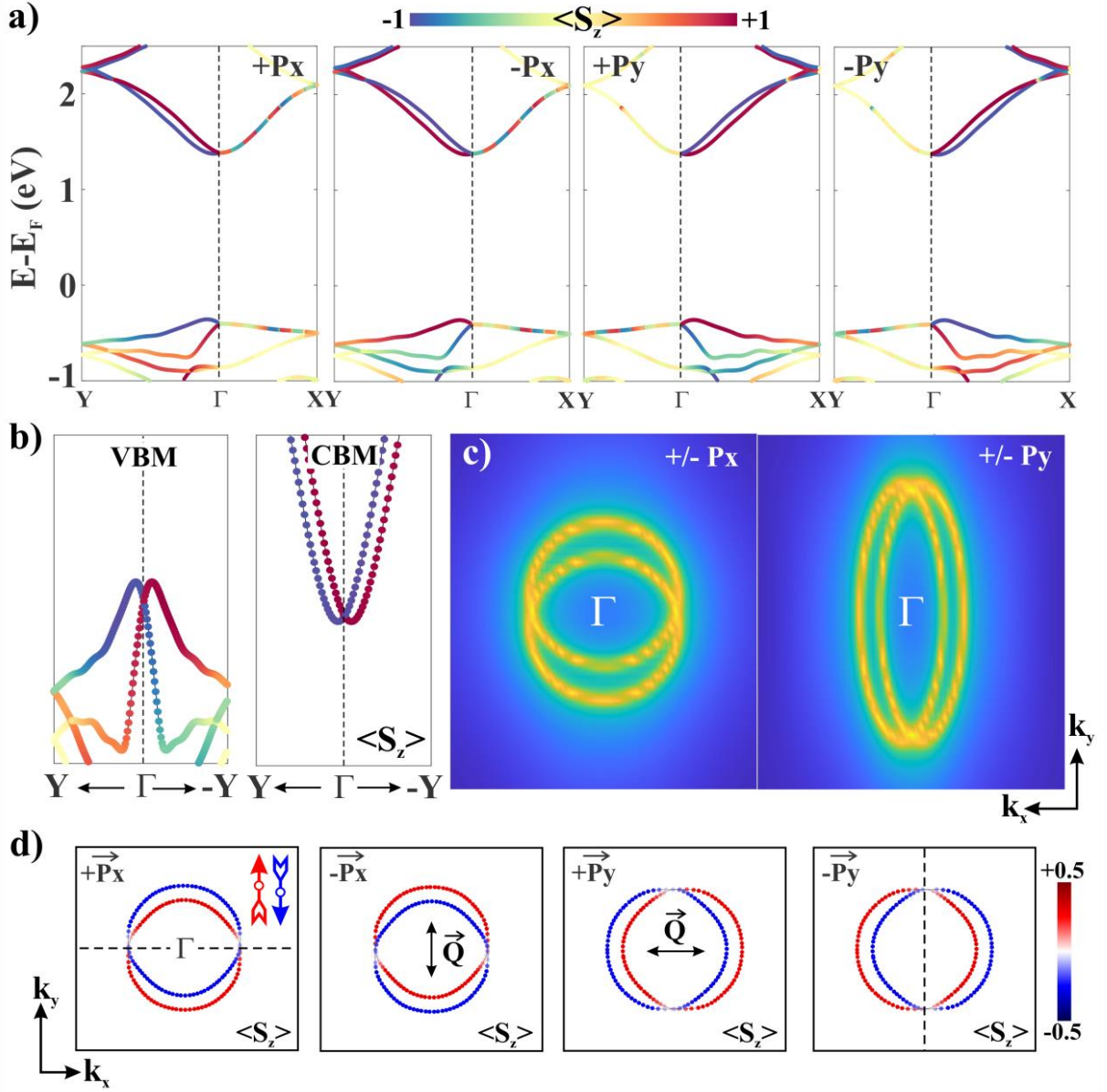

Figure 2. (a) spin-projected electronic band structure having ferroelectric polarization along $\pm P_x$ and $\pm P_y$ direction; the color represents the projection of $\langle S_z \rangle$ on each band, (b) a zoomed image of CBM and VBM showing out-of-plane Rashba splitting, (c) constant energy Fermi surfaces for ferroelectric polarization along $\pm P_x$ and $\pm P_y$ direction, (d) spin $\langle S_z \rangle$ projected constant energy 2D contour plots at $E-E_F = 1.9$ eV calculated in the $k_x$ - $k_y$ plane centred at the $\Gamma$-point; SDLN is indicated by a dashed line and the shifting wave vector is indicated by $\vec{Q}$.

## 2.3 Analytical Models in Spintronics and Persistent Spin Textures

To shed light on the origin of PST observed in this work and to determine the conditions for identification of an ideal PST material, understanding the spin textures proposed in the field of spintronics i.e., the difference between momentum-dependent and momentum-independent spin texture/ spin splitting is important. There are two primary models widely studied in the field of spintronics namely the Rashba and Dresselhaus models and two derived models namely the 2D quantum well and Dresselhaus [110] models which are less studied but relevant for designing a practical device.

**A. 2D Rashba and Dresselhaus Model**

The Rashba Hamiltonian is given by [29,30]: $\mathcal{H}_R = \frac{\hbar^2}{2m}(k_x^2 + k_y^2) + \alpha(\sigma_x k_y - \sigma_y k_x)$ (1)

$$\mathcal{H}_R = \begin{bmatrix} \frac{\hbar^2(k_x^2+k_y^2)}{2m} & (ik_x + k_y)\alpha \\ (-ik_x + k_y)\alpha & \frac{\hbar^2(k_x^2+k_y^2)}{2m} \end{bmatrix} \quad (2)$$

where $\vec{k}$ is the momentum of the electron, α corresponds to the strengths of Rashba spin-orbit coupling, m is the effective electron mass and $\sigma_i$ is Pauli spin matrices.

The energy dispersion can be obtained from the above Hamiltonian as;

$$E_{R_1/R_2} = \frac{\hbar^2(k_x^2+k_y^2)}{2m} \mp \alpha\sqrt{k_x^2 + k_y^2} \quad (3)$$

The spin polarization of each eigenstate can be obtained by $\vec{S} = \langle \psi_{\vec{k}} | \vec{\sigma} | \psi_{\vec{k}} \rangle$. The 3D view of energy dispersions, 2D contour plot, and spin textures is plotted in Figure 3 (a). The distinctive features of Rashba spin splitting, momentum-dependent and in-plane spin splitting can be observed where both inner and outer branches have opposite spin textures.

The Dresselhaus Hamiltonian is given by: $\mathcal{H}_D = \frac{\hbar^2}{2m}(k_x^2 + k_y^2) + \beta(\sigma_x k_x - \sigma_y k_y)$ (4)

$$\mathcal{H}_D = \begin{bmatrix} \frac{\hbar^2(k_x^2+k_y^2)}{2m} & (k_x + ik_y)\beta \\ (k_x - ik_y)\beta & \frac{\hbar^2(k_x^2+k_y^2)}{2m} \end{bmatrix} \quad (5)$$

where $\vec{k}$ is the momentum of the electron, $\beta$ corresponds to the strengths of Dresselhaus spin-orbit coupling, $m$ is the effective electron mass and $\sigma_i$ is Pauli spin matrices.

The energy eigenvalues and eigenstates ($\psi$) can be obtained by diagonalizing the Hamiltonian;

$$E_{D_1/D_2} = \frac{\hbar^2(k_x^2+k_y^2)}{2m} \mp \beta\sqrt{k_x^2 + k_y^2} \quad (6)$$

The 3D view of energy dispersions, 2D contour plots, and spin textures is plotted in Figure 3(b).

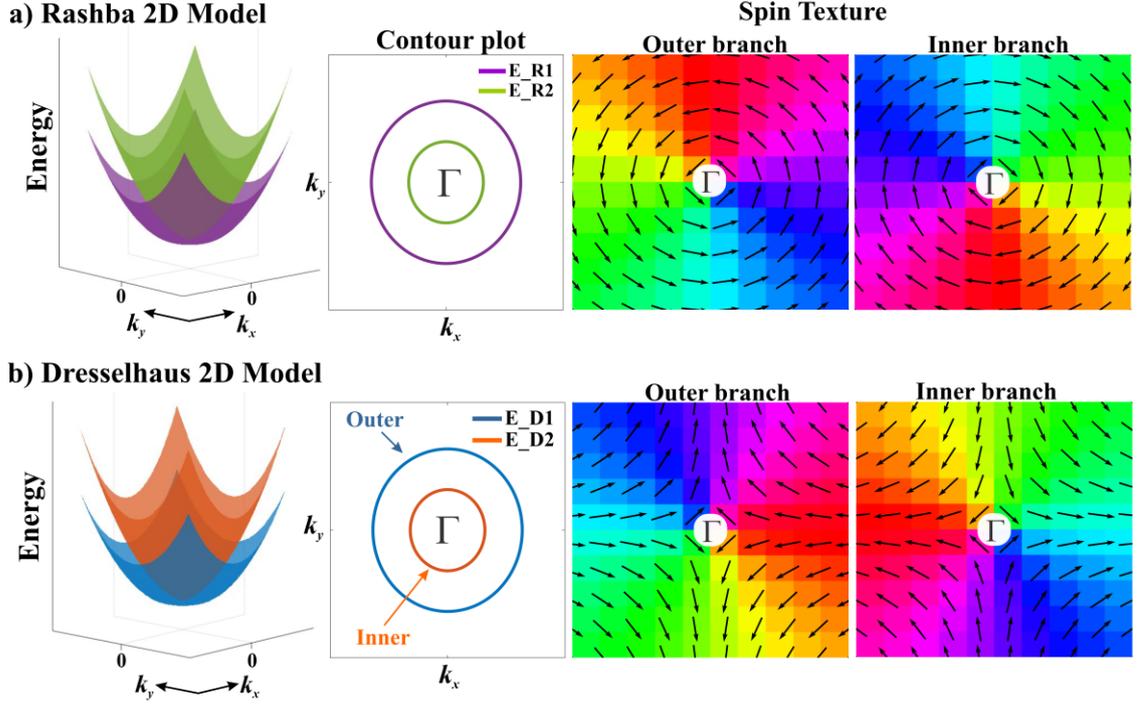

Figure 3. 3D plot of analytical energy dispersions, 2D constant energy contour plot, and spin textures for (a) Rashba and (b) Dresselhaus Hamiltonian. For plotting we have used $\hbar = m = \alpha = \beta = 1$, $k_x = k_y = [-1, 1]$; $\Gamma$ −point refers to (0,0) coordinate.

These primary two models show momentum-dependent spin texture. However, combining these two models can give a very distinctive spin texture under certain conditions as discussed below.

**B. 2D Quantum well model - Equal Rashba and Dresselhaus SOC Strength**

The linear terms of both Rashba and Dresselhaus contribution to the Hamiltonian for a 2D system are given by [4]:

$$\mathcal{H}_{RD} = \frac{\hbar^2 k^2}{2m} + \alpha(\sigma_x k_y - \sigma_y k_x) + \beta(\sigma_x k_x - \sigma_y k_y) \qquad (7)$$

The Hamiltonian takes the form;

$$\mathcal{H}_{RD} = \begin{bmatrix} \frac{(k_x^2 + k_y^2)\hbar^2}{2m} & (ik_x + k_y)\alpha + (k_x + ik_y)\beta \\ (-ik_x + k_y)\alpha + (k_x - ik_y)\beta & \frac{(k_x^2 + k_y^2)\hbar^2}{2m} \end{bmatrix} \qquad (8)$$

Diagonalizing the Hamiltonian in equation (8) gives the energy eigenvalues as;

$$E_\pm = \frac{\hbar^2(k_x^2 + k_y^2)}{2m} \pm \sqrt{k_x^2 \alpha^2 + k_y^2 \alpha^2 + 4k_x k_y \alpha\beta + k_x^2 \beta^2 + k_y^2 \beta^2} \qquad (9)$$

Now simplifying the Hamiltonian with equal Rashba and Dresselhaus strength, the energy eigenvalues and corresponding energy states have been calculated and provided in Table 1 and illustrated in Figure 4. For equal Rashba and Dresselhaus strength $\alpha = \pm\beta$, a spin degenerate line node and the shifting wave vector $\vec{Q} = \frac{\sqrt{2}m\alpha}{\hbar^2}(1, \pm 1)$ between spin-up and spin-down parabolic bands can be observed which is consistent with the literature. Only, under the condition of $\alpha = \pm\beta$, the spin state of the electrons become independent of the wave vector $k$ which can be depicted from the spin textures of $A_1/A_2$ and $A_3/A_4$ plotted in Figure 4(a) whereas for $\alpha \neq \beta$ the spin textures become momentum dependent as can be observed in Figure 4(b). In the experiment, $\alpha$ and $\beta$ can be tuned with an external electric field and sample geometry.

| Table 1 The energy eigenvalues and corresponding eigenstates of Hamiltonian having equal Rashba and Dresselhaus strength ||| 
|---|---|---|
| Cases | Eigenvalues | Eigenstates |
| $\alpha = +\beta$ | $E_{A1} = \frac{\hbar^2(k_x^2 + k_y^2)}{2m} - \sqrt{2}\alpha(k_x + k_y)$ | $\begin{pmatrix} -\frac{(1+i)(k_x + k_y)}{\sqrt{2}} \\ 1 \end{pmatrix}$ |
|  | $E_{A2} = \frac{\hbar^2(k_x^2 + k_y^2)}{2m} + \sqrt{2}\alpha(k_x + k_y)$ | $\begin{pmatrix} \frac{(1+i)(k_x + k_y)}{\sqrt{2}} \\ 1 \end{pmatrix}$ |
| $\alpha = -\beta$ | $E_{A3} = \frac{\hbar^2(k_x^2 + k_y^2)}{2m} - \sqrt{2}\alpha(k_x - k_y)$ | $\begin{pmatrix} \frac{(1-i)(k_x - k_y)}{\sqrt{2}} \\ 1 \end{pmatrix}$ |
|  | $E_{A4} = \frac{\hbar^2(k_x^2 + k_y^2)}{2m} + \sqrt{2}\alpha(k_x - k_y)$ | $\begin{pmatrix} -\frac{(1-i)(k_x - k_y)}{\sqrt{2}} \\ 1 \end{pmatrix}$ |

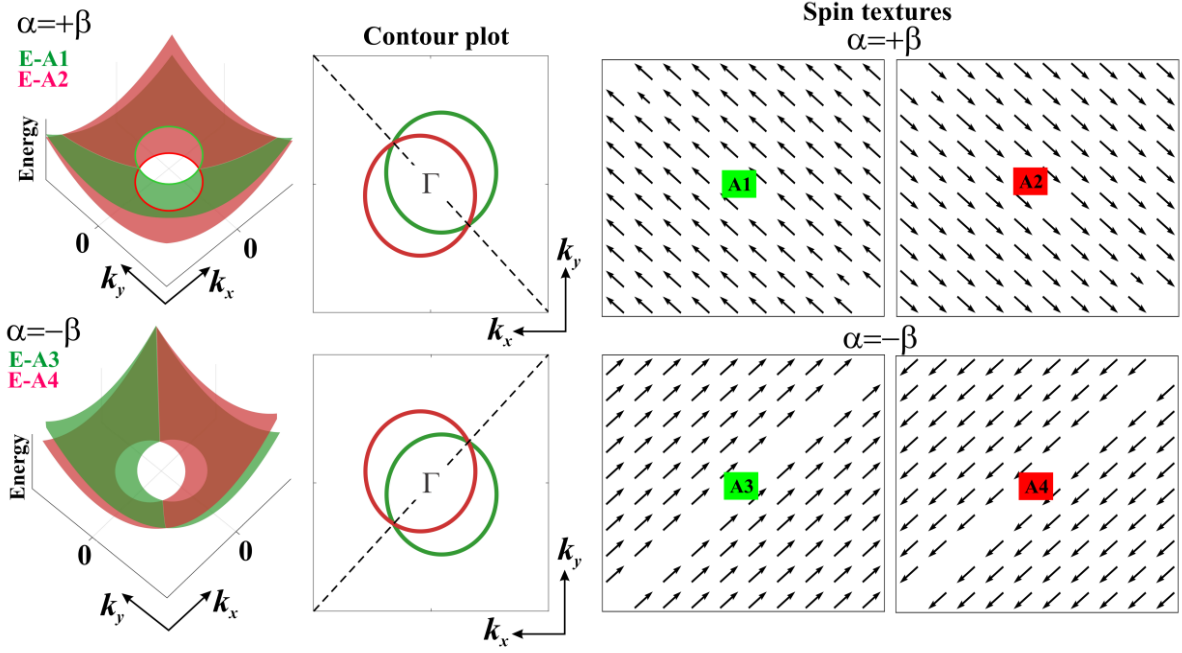
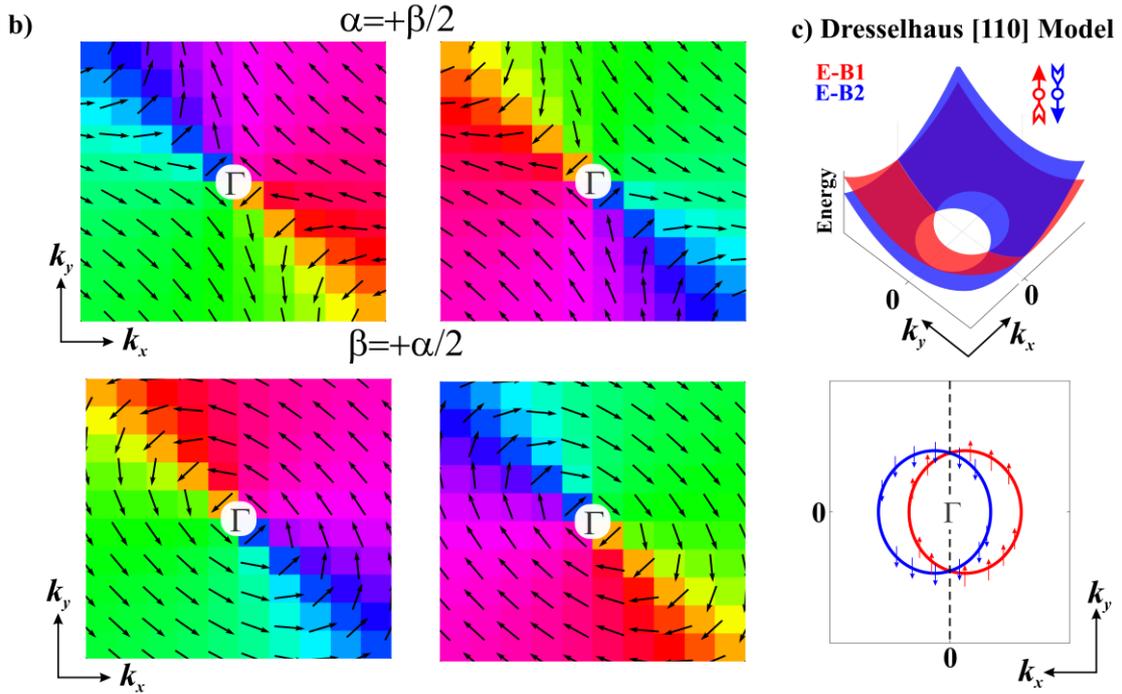

Figure 4. (a) 3D plot of analytical energy dispersions, 2D constant energy contour plot, spin textures obtained from Hamiltonian $\mathcal{H}_{RD}$ for equal Rashba and Dresselhaus strength ($\alpha = \beta$), (b) spin textures for $\alpha \neq \beta$, (c) spin projected 3D energy dispersion and 2D energy contour plots obtained from Dresselhaus [110] model.

## C. Dresselhaus [110] Model

The diagonalized Hamiltonian presented in equation (8) for equal strength of Rashba and Dresselhaus SOC is mathematically equivalent to the Hamiltonian of [110] Dresselhaus model and is given by;

$$\mathcal{H}_{[110]} = \frac{k_x^2 + k_y^2}{2m} - 2\alpha k_x \sigma_z \qquad (10)$$

The energy eigenvalues, eigenvectors and spin polarizations are listed in Table 2 and illustrated in Figure 4(c).

| Table 2 The energy eigenvalues, eigenstates, and corresponding spin polarization for the Dresselhaus [110] model | | |
|---|---|---|
| Eigenvalues | Eigenvectors | Spin polarization in the form of $\begin{pmatrix} S_x \\ S_y \\ S_z \end{pmatrix}$ |
| $E_{B1} = \dfrac{\hbar^2(k_x^2 + k_y^2)}{2m} - 2\alpha k_x$ | $\begin{pmatrix} 1 \\ 0 \end{pmatrix}$ | $\begin{pmatrix} 0 \\ 0 \\ 1 \end{pmatrix}$ |
| $E_{B2} = \dfrac{\hbar^2(k_x^2 + k_y^2)}{2m} + 2\alpha k_x$ | $\begin{pmatrix} 0 \\ 1 \end{pmatrix}$ | $\begin{pmatrix} 0 \\ 0 \\ -1 \end{pmatrix}$ |

In MgTe (110) facets, the in-plane ferroelectric polarization $P_x$ can generate a unidirectional spin-orbit-field (SOF) in a 2D material expressed by;

$$\vec{\Omega}_{SOF}(\vec{k}) = \alpha(\hat{P}_x \times \vec{k}) = \alpha k_y \hat{z} \qquad (11)$$

Now, considering the direction of polarization along the +x axis i.e., $+\hat{P}_x$, the effective Hamiltonian including the SOC term can be written as:

$$\mathcal{H} = \mathcal{H}_{kin} + \vec{\Omega}_{SOF} \cdot \vec{\sigma} = \frac{\hbar^2}{2m}(k_x^2 + k_y^2) + \alpha k_y \sigma_z \qquad (12)$$

which takes an identical form to that of Dresselhaus [110] model as in equation (10).

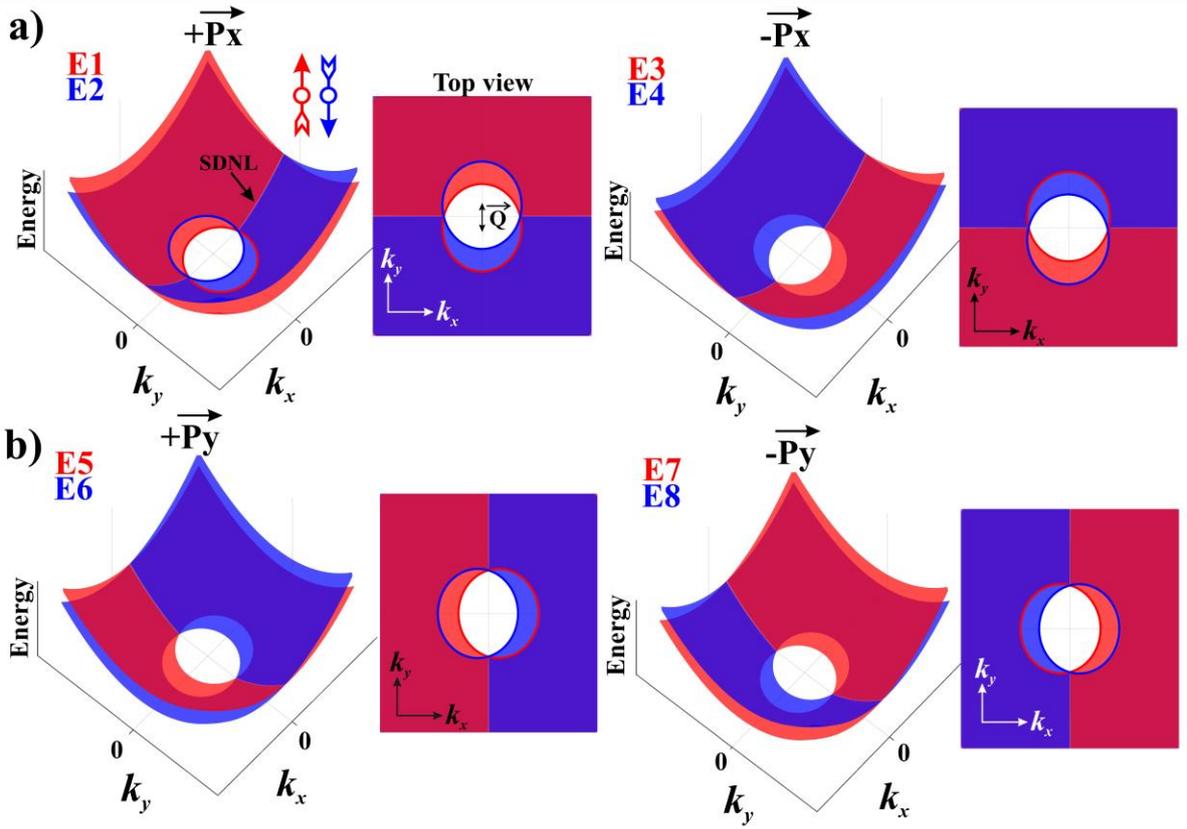

Figure 5. 3D view of energy dispersion obtained by diagonalizing Hamiltonians listed in Table-1; spin degenerate line node (SDLN) and shift vector $\vec{Q}$ are indicated. Here $\hbar = m = 1, \alpha = 0.2, k_x = k_y = [-\pi, +\pi]$ are used for plotting.

For a systematic analysis, the energy relations, eigenstates, and corresponding spin polarization for different ferroelectric polarization directions $\pm \vec{P}_x, \pm \vec{P}_y$ are listed in Table 3 and plotted in Figure 5. For a given energy dispersion, the eigenstates/spin polarizations are completely reversed with ferroelectric polarization which can also be seen from the colours in Figure 5. These analytical results obtained in this section match exactly with those calculated using DFT in Figure 2(d).

| Table 3 The energy eigenvalues, eigenstates, and corresponding spin polarization for a 2D system having ferroelectric polarization direction along $+\widehat{P}_y$ |||
|---|---|---|
| Energy eigenvalues | Eigenstates | Spin polarization in the form of $\begin{pmatrix} S_x \\ S_y \\ S_z \end{pmatrix}$ |
| $+\widehat{P}_x$: $\mathcal{H}_1 = \dfrac{\hbar^2}{2m}(k_x^2 + k_y^2) + \alpha k_y \sigma_z$ |||
| $E_1 = \dfrac{\hbar^2(k_x^2 + k_y^2)}{2m} + \alpha k_y$ | $\begin{pmatrix} 1 \\ 0 \end{pmatrix}$ | $\begin{pmatrix} 0 \\ 0 \\ 1 \end{pmatrix}$ |
| $E_2 = \dfrac{\hbar^2(k_x^2 + k_y^2)}{2m} - \alpha k_y$ | $\begin{pmatrix}0\\1\end{pmatrix}$ | $\begin{pmatrix}0\\0\\-1\end{pmatrix}$ |
| $-\hat{P}_x$: $\mathcal{H}_2 = \dfrac{\hbar^2}{2m}(k_x^2 + k_y^2) - \alpha k_y \sigma_z$ | | |
| $E_2 = \dfrac{\hbar^2(k_x^2 + k_y^2)}{2m} - \alpha k_y$ | $\begin{pmatrix}1\\0\end{pmatrix}$ | $\begin{pmatrix}0\\0\\1\end{pmatrix}$ |
| $E_3 = \dfrac{\hbar^2(k_x^2 + k_y^2)}{2m} + \alpha k_y$ | $\begin{pmatrix}0\\1\end{pmatrix}$ | $\begin{pmatrix}0\\0\\-1\end{pmatrix}$ |
| $+\hat{P}_y$: $\mathcal{H}_3 = \dfrac{\hbar^2}{2m}(k_x^2 + k_y^2) - \alpha k_x \sigma_z$ | | |
| $E_4 = \dfrac{\hbar^2(k_x^2 + k_y^2)}{2m} - \alpha k_x$ | $\begin{pmatrix}1\\0\end{pmatrix}$ | $\begin{pmatrix}0\\0\\1\end{pmatrix}$ |
| $E_5 = \dfrac{\hbar^2(k_x^2 + k_y^2)}{2m} + \alpha k_x$ | $\begin{pmatrix}0\\1\end{pmatrix}$ | $\begin{pmatrix}0\\0\\-1\end{pmatrix}$ |
| $-\hat{P}_y$: $\mathcal{H}_4 = \dfrac{\hbar^2}{2m}(k_x^2 + k_y^2) + \alpha k_x \sigma_z$ | | |
| $E_6 = \dfrac{\hbar^2(k_x^2 + k_y^2)}{2m} + \alpha k_x$ | $\begin{pmatrix}1\\0\end{pmatrix}$ | $\begin{pmatrix}0\\0\\1\end{pmatrix}$ |
| $E_7 = \dfrac{\hbar^2(k_x^2 + k_y^2)}{2m} - \alpha k_x$ | $\begin{pmatrix}0\\1\end{pmatrix}$ | $\begin{pmatrix}0\\0\\-1\end{pmatrix}$ |

Considering the symmetry of MgTe(110) 2D crystal structure, including a two-fold rotational symmetry $C_{2x}$ around the x-axis, time-reversal symmetry, and mirror symmetry in y ($M_y$), a *k.p* effective Hamiltonian [31] can be constructed as:

$$\mathcal{H}_5 = \frac{\hbar^2}{2m}(k_x^2 + k_y^2) + \varphi k_y \sigma_z \tag{13}$$

Also, an equivalent Hamiltonian expression can be obtained by considering a two-fold rotational symmetry $C_{2y}$ around the *y*-axis, time-reversal symmetry, and mirror symmetry in *x* ($M_x$) given by;

$$\mathcal{H}_6 = \frac{\hbar^2}{2m}(k_x^2 + k_y^2) + \varphi k_x \sigma_z \tag{14}$$

This section gives a broad overview of the differences between momentum-dependent and momentum-independent spin textures/splitting which can be depicted by comparing the spin textures plotted in Figure 3 (2D Rashba/ Dresselhaus models) and Figure 4 (2D quantum well model under the condition of equal Rashba and Dresselhaus SOC/ Dresselhaus [110] model). The derived quantum well model proposed a new route to obtain a momentum-independent unidirectional spin polarization under stringent conditions of equal Rashba and Dresselhaus SOC strength but impractical for realizing the actual device. Further, analyzing these previously proposed theoretical models, the characteristic properties of an ideal PST material can be of the followings;

(i) An ideal material should exhibit PST at the Brillouin zone centre i.e., at the $\Gamma$−point.
(ii) Like Rashba semiconductors exhibit characteristic concentric Fermi surfaces, PST material should exhibit the distinctive constant energy Fermi surfaces as plotted in Figure 4.
(iii) The bands in PST material should satisfy the "one-band one spin polarization state" as observed in Figure 4 and both bands should have opposite spin polarizations.

All these electronic features listed are similar to that of the Rashba semiconductor. [32] Is there any material that shows PST intrinsically and exhibits characteristic electronic/spintronic features as mentioned? Although there are reports of observation of PST in a few materials, none of them are ideal. Bulk BiInO$_3$ proposed by Tao and Tsymbal shows PST at X (0.5,0) and Y (0,0.5) high symmetry point of Brillouin zones (BZ) which results in partial PST as mentioned in their report [14]; here partial PST refers to unidirectional spin polarization in a partial region of BZ rather than full BZ. All the family members of 2D SnS, SnSe, SnTe, GeS, GeSe, and GeTe exhibit PST at X (0.5,0) or Y (0, 0.5) points of 2D orthorhombic BZ and hence exhibit partial PST. Although MgTe (110) facets have similar crystal structures to that of GeTe and SnTe, the facets proposed in this work exhibit significantly different electronic/spintronic properties. For a systematic comparison, the band structure with SOC and spin texture of band edges are plotted in Figure 6.

The spin projected band structure of GeTe in Figure 6(a) indicates bands along X→S are spin degenerate. Now mapping the spin textures of the two highest energy valence bands i.e., VBM and VBM-1 bands, each bands have both spin polarization i.e., $\pm S_z$ which violates the "one-band one spin polarization" condition. The spin texture is complicated as the spin direction abruptly changes its sign at $k_y$=0. A similar spin texture has been earlier observed in the case of bulk BiInO$_3$. Such behavior of spin cannot be explained using Dresselhaus [110] Hamiltonian but requires a complex Hamiltonian similar to that proposed by Tao and Tsymbal. In contrast to these existing reports of partial PST materials, MgTe (110) shows momentum-independent unidirectional spin polarization in the entire Brillouin zone which can be depicted in Figure 6(b). The spin textures of the lowest two conduction bands i.e., CBM and CBM+1 are plotted in Figure S5. The Fermi surfaces of MgTe (110) facets plotted in Figure 2 match exactly with that of the characteristics Fermi surface of PST material as obtained from the Hamiltonians 2D quantum well model or Dresselhaus [110] model as in Figure 4. Moreover, MgTe (110) facets satisfy all three requirements mentioned earlier and hence an ideal PST material.

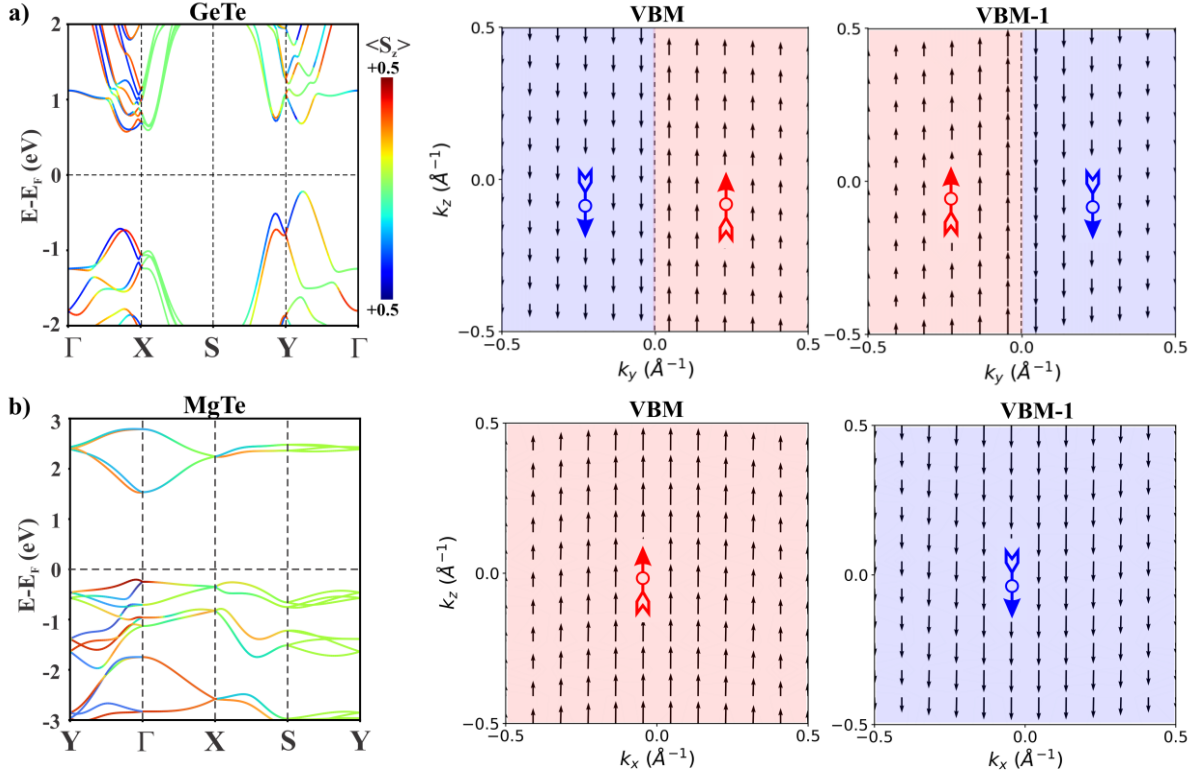

Figure 6 DFT obtained electronic band structure including SOC and spin textures of two highest energy valence bands VBM and VBM-1 of (a) GeTe and (b) MgTe.

## 2.4 Magnetic Element-Free Nonballistic Spin-Field Effect Transistor

To circumvent the existing limitations in conventional s-FET, a schematic of an all-electric spin-Hall transistor (SHT) [33] is presented in Figure 7. The operation of such devices relies on different mechanisms such as the spin-Hall effect (SHE) and inverse spin-Hall effect (ISHE). The SHT consists of only one material with three regions. The direct SHE converts the charge current flowing along the y-direction to a transverse spin-current along the x-direction where the spin polarization is along the z-axis given by; $I_s \propto \vartheta \times I_c$. Here, $\vartheta$ is the spin polarization direction and $I_s/I_c$ is the flow of spin/charge current. Thus, the pure spin current moving along the x-axis is injected into region M-II. The spins polarized along the z-axis are efficiently injected from M-I to M-II without loss. The spin does not lose any information for $E_z = 0$. However, previous reports indicate the breaking of PSH mode under finite vertical electric field. The gate-modulated spins in M-II are now injected into the region M-III where the ISHE converts the spin-current +x into a transverse charge current -y given by $I_c \propto \vartheta \times I_s$ which generates a Hall voltage ($V_H$) along the y-direction. [34] Unlike conventional s-FET, no interface between regions M-II and M-III guarantees an efficient detection of spin information. These three processes provide an essential platform to generate, manipulate, and detect the spins without involving ferromagnets or external magnetic fields. Thus, the Hall voltage in M-III can be controlled in region M-II with an out-of-plane electric field. [35–37] The proposed device model in this work is inspired by recent experimental work [9] where the spins polarized along the z-axis are generated using the spin-Hall effect. This device model is different from the earlier quantum well model [5] where a constant electric field is applied to sustain PST which cannot be an energy-efficient model. Also, our proposed device model needs to be modified for spin polarization along the *x* or *y* direction as predicted in bulk nonsymmorphic structures. Since 2L-MgTe is the basic building block, the coordination of the $Mg^{2+}$ ion is three-fold whereas four-fold in the 3L-MgTe structure. To show the independence of PST with layer thickness and robustness against mechanical strain, the spin-projected band structure of 3L MgTe and strained 2L-MgTe are plotted in Figure S6-S7 which confirms the same. The Rashba constant for 3L MgTe is calculated to be $\alpha_R$= 0.55 eV/Å and the pitch of the PSH $l_{PSH}$ is calculated to be 10.6 nm. The spin-Hall conductivity of 2L and 3L MgTe systems are calculated and plotted in Figure 7. The spin Hall conductivity (SHC) is a measure of the spin Hall effect (SHE) which originates from the coupling between charge and spin degrees of freedom due to spin-orbit interaction. [38,39] The SOC transforms longitudinal charge flow into transverse spin current without any external magnetic field. The SHC projected band structure and *k*-point resolved SHC are plotted in Figure 7(b) which indicates a finite SHC observed at both CBM and VBM. The SHC is found to be concentrated around the Γ -point of the rectangular Brillouin zone. [40]

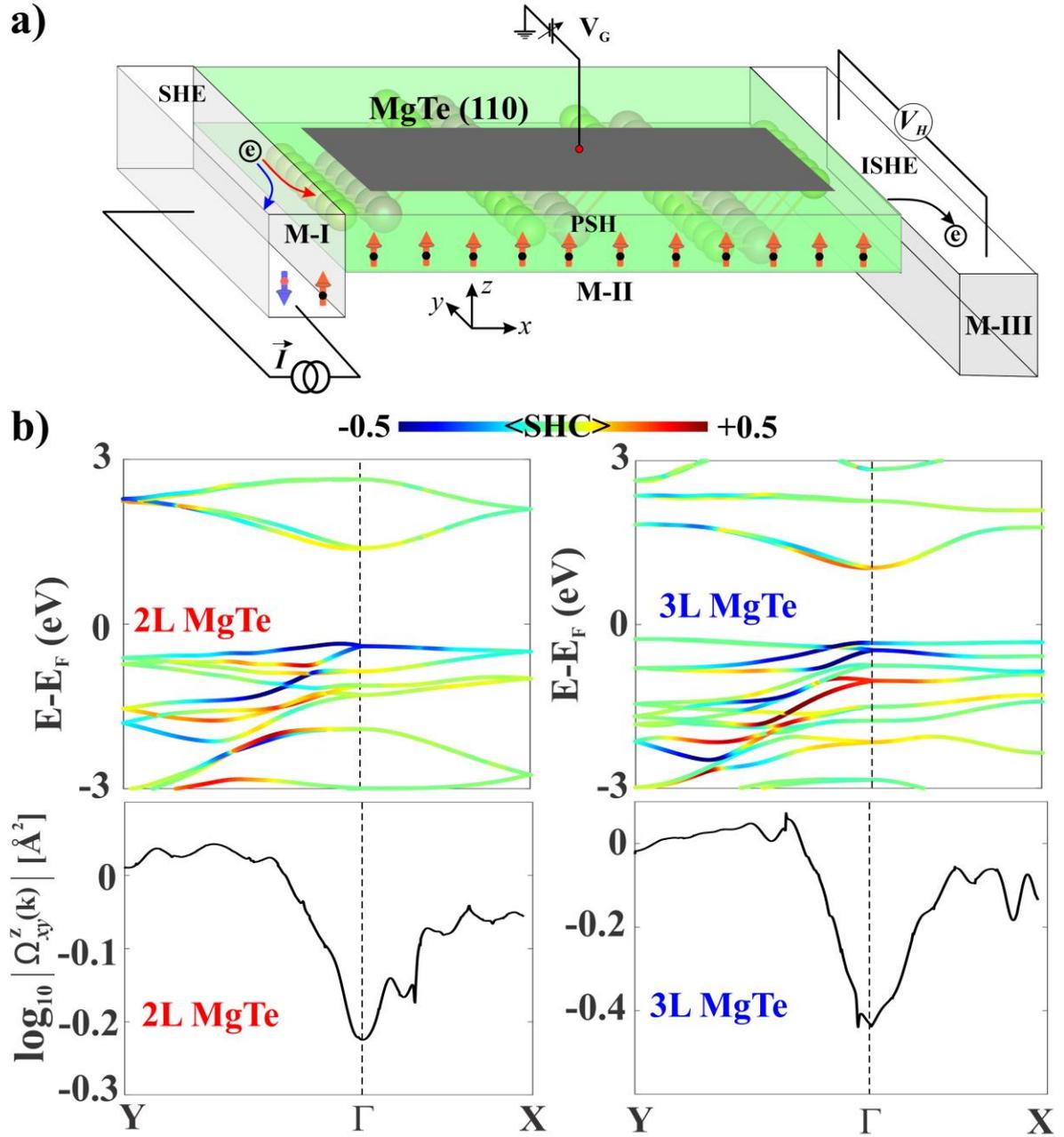

Figure 7. (a) Schematic of spin-Hall effect transistor based on (110) facets of MgTe, (b) SHC projected band structure, and *k*-point resolved SHC for 2L and 3L MgTe.

**Conclusion**

In conclusion, (110) facets of zinc-blende MgTe 2D structures possess a unique combination of the three basic transformations found in nature; rotation, mirror reflection, and translation, and thus exhibit uncommon properties with respect to its parent bulk structure or other facets. The in-plane ferroelectric polarization with SOC ushers unconventional momentum-independent unidirectional spin polarization in the full Brillouin zone. Based on the results from previously proposed analytical models, a list of characteristic electronic properties are listed to identify an ideal PST material. A systematic and comparative approach revealed the uniqueness of MgTe (110) facets compared to that of other reports such as GeTe which exhibit partial PST. To understand the origin of such exquisite properties in this particular facet, density functional theory is employed and $k.p$ effective Hamiltonian is constructed using the method of invariance and their results are compared. MgTe (110) facets satisfy all the listed conditions for an ideal PST material and the results from the DFT and analytical model exactly matched. Further, 2L and 3L-MgTe (110) facets exhibit finite spin Hall conductivity at the band edges and are mostly localized at the Γ-point. Although the MgTe (110) has been studied for many years in optoelectronic devices, this work sheds light on its practical application in nonballistic spin-field effect transistors. The future of spintronics lies in the realization of ferromagnet-free nonballistic s-FET using 2D semiconductors that exhibit intrinsic symmetry-protected PST intrinsically such as MgTe (110) proposed in this work.

## 4. Computational details:

All the theoretical calculations are performed using density functional theory (DFT) embedded in the VASP code [41,42] with the Perdew-Burke-Ernzerhof generalized gradient approximation (GGA-PBE). [43] The projector-augmented plane waves (PAW) have been adopted to describe the ion-electron interaction. [44] The plane wave cut-off of 520 eV is used with the energy convergence criteria of $10^{-8}$ eV. A Γ-centered $k$-point mesh of 15×15×1 is used to sample the Brillouin zone (BZ). A vacuum space of more than 20Å is added to avoid periodic interaction along the $z$-direction. PYPROCAR is used to plot spin texture. [45] The ferroelectric properties are calculated using the Berry phase method. [46–48] The spin Hall conductivity (SHC) is calculated from the Berry curvature using a dense 150×150×1 $k$-mesh based on maximally localized Wannier function embedded in Quantum Espresso [49,50] and Wannier90 code [51,52] by the formula [53,54];

$$\sigma_{xy}^z = \frac{e}{\hbar} \int_{Bz} \frac{dk}{(2\pi)^2} \Omega^z(k)$$

$\Omega^z(k)$ is the k-resolved term which is given by $\Omega^z(k) = \sum_n f_{kn} \Omega_n^z(k)$. Here, $f_{kn}$ is the Fermi-Dirac distribution function for the $n$th band at k and $\Omega_n^z(k)$ is an analogue of the Berry curvature for the $n$th band given as;

$$\Omega_n^z(k) = \sum_{n' \neq n} \frac{2Im[\langle kn|j_x^z|kn'\rangle \langle kn'|v_y|kn\rangle]}{(\epsilon_{kn} - \epsilon_{kn'})^2}$$

Here, $j_x^z = \frac{1}{2}\{s_z, v\}$ is the spin current operator, $s_z = \frac{\hbar}{2}\sigma^z$ is the spin operator, $v$ is the velocity operator and $|kn\rangle$ is the wave function of energy $\epsilon_{kn}$.

The charge carrier mobility is calculated using the ab initio Boltzmann transport equation in the framework of the self-energy relaxation time approximation as implemented in PERTURBO. [55] The dynamical matrix is computed on a 8×8×1 $q$-point mesh in the phonon calculations. The phonon modes and frequencies at other general $k$-points are then computed by Fourier transformation of the dynamical matrix in reciprocal space. The relaxation times arising from electron-phonon scattering are calculated using the PERTURBO package which utilizes the Wannier interpolation scheme. PERTURBO interpolates the electron-phonon coupling matrices as well as electron and phonon eigenvalues from a coarse grid to a fine grid 120×120×1. The phonon dispersions are obtained using finite displacement method encoded in the PHONOPY code [56] using a $4 \times 4 \times 1$ supercell.

**Acknowledgment**

MKM and PJ acknowledge financial support from the U.S. Department of Energy, Office of Basic Energy Sciences, Division of Materials Sciences and Engineering under Award No. DE-FG02-96ER45579. Resources of the National Energy Research Scientific Computing (NERSC) Center supported by the Office of Science of the U.S. Department of Energy under Contract No. DE-AC02-05CH11231 is also acknowledged. The authors extend their acknowledgment to the High-Performance Research Computing (HPRC) core facility at Virginia Commonwealth University for providing supercomputing resources.